# Strong Quantum Confinement Effects and Chiral Excitons in Bio-Inspired ZnO-Amino Acid Co-Crystals

*Madathumpady Abubaker Habeeb Muhammed,*[1,2] *Marlene Lamers,*[1,2] *Verena Baumann,*[1,2] *Priyanka Dey,*[1,2] *Adam J. Blanch,*[1,2] *Iryna Polishchuk,*[3] *Xiang-Tian Kong,*[4,5] *Davide Levy,*[3] *Alexander Urban,*[1,2] *Alexander O. Govorov,*[4*] *Boaz Pokroy,*[3*] *Jessica Rodríguez-Fernández,*[1,2]*

*Jochen Feldmann*[1,2]

[1]Department of Physics and Center for NanoScience (CeNS), Ludwig-Maximilians-Universität München, Amalienstr. 54, 80799 Munich (Germany)

[2]Nanosystems Initiative Munich (NIM), Schellingstr. 4, 80799 Munich (Germany)

[3]Department of Materials Science and Engineering and the Russell Berrie Nanotechnology Institute, Technion — Israel Institute of Technology, 32000 Haifa, Israel

[4]Department of Physics and Astronomy, Ohio University, OH 45701, Athens, USA

[5]Institute of Fundamental and Frontier Sciences and State Key Laboratory of Electronic Thin Films and Integrated Devices, University of Electronic Science and Technology of China, Chengdu 610054, China

ABSTRACT

Elucidating the underlying principles behind band gap engineering is paramount for the successful implementation of semiconductors in photonic and optoelectronic devices. Recently it has been shown that the band gap of a wide and direct band gap semiconductor, such as ZnO, can be modified upon co-crystallization with amino acids, with the role of the biomolecules remaining unclear. Here, by probing and modeling the light emitting properties of ZnO-amino acid co-crystals, we identify the amino acids' role on this band gap modulation and demonstrate their effective chirality transfer to the inter-band excitations in ZnO. Our 3D quantum model suggests that the strong band edge emission blue shift in the co-crystals can be explained by a quasi-periodic distribution of amino acid potential barriers within the ZnO crystal lattice. Overall, our findings indicate that biomolecule co-crystallization can be used as a truly bio-inspired means to induce chiral quantum confinement effects in quasi-bulk semiconductors.

KEYWORDS

Semiconductors, zinc oxide, biomolecules, amino acids, quantum confinement, band gap, chirality

TOC GRAPH

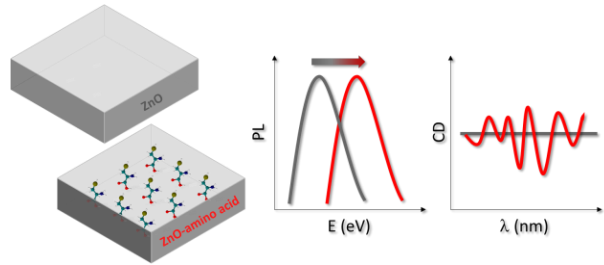

The ability to on-demand modify the band gap of wide and direct band gap semiconductors has thrilled researchers for decades prompted by their prospects in photonic and optoelectronic devices. ZnO has been one of the most investigated wide and direct band gap semiconductors. With a band gap at *ca.* 3.3 eV, ZnO is ideally suited for lighting applications.[1] Also its high exciton binding energy (*ca.* 60 meV) offers a competitive advantage with respect to other direct gap semiconductors, such as GaN, as it enables device operation at considerably higher temperatures due to the effective prevention of exciton dissociation.[1-2] The band gap of ZnO can be conveniently altered without posing changes in the chemical composition of the semiconductor. For instance, both theoretical (*e.g.*, ab initio density functional theory)[3] and experimental[4-8] works have demonstrated that tensile, compressive or bending strain can be used as an effective means to tune the band gap of ZnO. The strain induces changes in the position of the Zn and O atoms within the crystal lattice that lead to changes in electron density and charge distribution resulting in a corresponding band gap change.[3] Miniaturization is another approach for engineering the band gap of ZnO without altering its chemical composition. As it occurs with other semiconductors,[9-10] when the dimensions of ZnO crystals approach the exciton Bohr radius their electronic properties start to change, giving rise to strong changes in their band gap as a result of quantum size effects.[11-14] The band gap of ZnO can be controllably engineered by means of chemical doping as well. Here, the introduction of foreign atoms leads to band gap modifications along with compositional changes. In this regard, prominent band gap modifications have been reported for doping with Ni,[15] Al,[16-17] or Mg,[18] but also for alloying with MgO and CdO.[19-25]

A leap in band gap engineering has recently emerged as a result of applying biomimetic principles to semiconductor materials.[26] Pokroy and co-workers demonstrated that, similarly to $CaCO_3$ minerals[27] and biomimetic calcite,[28-29] a semiconductor like ZnO can also undergo co-crystallization with biomolecules, such as amino acids, leading to strong band gap changes.[26] Those band gap modifications were tentatively attributed to the increasing lattice distortions resulting from the anisotropic tensile strain along the *a* and *c* axes that occurs upon amino acid co-crystallization.[26, 30] However, even though the average tensile strain in ZnO-amino acid co-crystals is on average *ca.* 10-fold smaller than previously investigated samples,[4] the change in the co-crystals' band gap is significantly larger and, remarkably, of a different polarity. Therefore, since the band gap changes in ZnO-amino acid co-crystals cannot be unambiguously attributed to lattice distortions, it remains unclear how the amino acids modify the band gap. In this work we shed light onto the actual role that the amino acids play on the strong band gap modulation of ZnO-amino acid co-crystals by probing and modeling their light emission properties. The good agreement between the experimental and calculated band edge emission blue shifts indicate that the co-crystals can be conceived as a 3D superlattice of quasi-periodically-arranged amino acids within a ZnO matrix. The biomolecules serve as potential barriers for electrons and holes, favoring their radiative recombination and leading to strong quantum confinement effects that account for the observed increases in band gap. Furthermore, through circular dichroism analysis we demonstrate that the co-crystals display an effective chirality transfer from the amino acids to the inter-band excitations in ZnO. Therefore, our work points to semiconductor-biomolecule co-crystals as unique quantum-confined materials with intrinsic chiral excitons.

With the aim of shedding light onto the role of co-crystallized amino acids on the light-matter interactions of ZnO-co-crystals, we synthesized three representative ZnO crystal types (ZnO-Ref, ZnO-Tyr, and ZnO-Cys)[i] as reported by Brif *et al.*[26] (see Experimental Section in the Supporting Information). Pure ZnO crystals (ZnO-Ref hereafter) were prepared in the absence of co-crystallizing amino acids and featured dimensions of ~ 0.5-2.5 μm along with a flower-like morphology (see SEM micrographs in **Figure 1**a). Co-crystals synthesized in the presence of tyrosine (ZnO-Tyr) featured a similar size and morphology (**Figure 1**b), while those obtained in the presence of co-crystallizing cysteine (ZnO-Cys) were less homogeneous in size and shape (**Figure 1**c). The presence of the amino acids in the co-crystals was confirmed by wavelength-dispersive X-ray spectroscopy (WDS) analysis, with the specific atomic percentages (at.%) of Zn and N (the amino acid-relevant atom) being, respectively: $57.00 \pm 0.37$ and $0.19 \pm 0.03$ for ZnO-Tyr and $54.02 \pm 0.46$ and $1.70 \pm 0.03$, for ZnO-Cys. The synchrotron high-resolution XRD patterns (**Figure 1**d) of all three samples confirmed their wurtzite crystalline structure, with only a minor impurity phase in the case of ZnO-Cys (Figure S1), and the significant anisotropic lattice distortion in the co-crystals with respect to ZnO-Ref, namely: 0.1124 % for ZnO-Tyr *vs.* 0.1431 % for ZnO-Cys along the *a* axis and 0.1304 % for ZnO-Tyr *vs.* 0.1325 % for ZnO-Cys along the *c* axis (see Rietveld refinement plots in Figure S2 and their quantitative results in **Table 1**). The crystallinity of the samples was confirmed, as shown as an example for the ZnO-Cys sample, *via* XRD analysis by using an internal alumina standard both before (Figure S3) and after heating (Figure S4).

---

[i] After an initial screening of the photoluminescence and circular dichroism of several ZnO-amino acid samples, we found that these samples (ZnO-Tyr and ZnO-Cys) were the ones featuring a better-defined band edge emission and a sufficiently intense CD signal. For this reason we decided to focus our optical studies on these representative ZnO-amino acid samples. The results obtained in this work for those ZnO-amino acid co-crystals (namely, the band edge emission increase upon amino acid incorporation, the proposed quantum model to explain such a band edge emission increase, and the chiral excitons/defects) can be extrapolated to other ZnO-amino acid co-crystals.

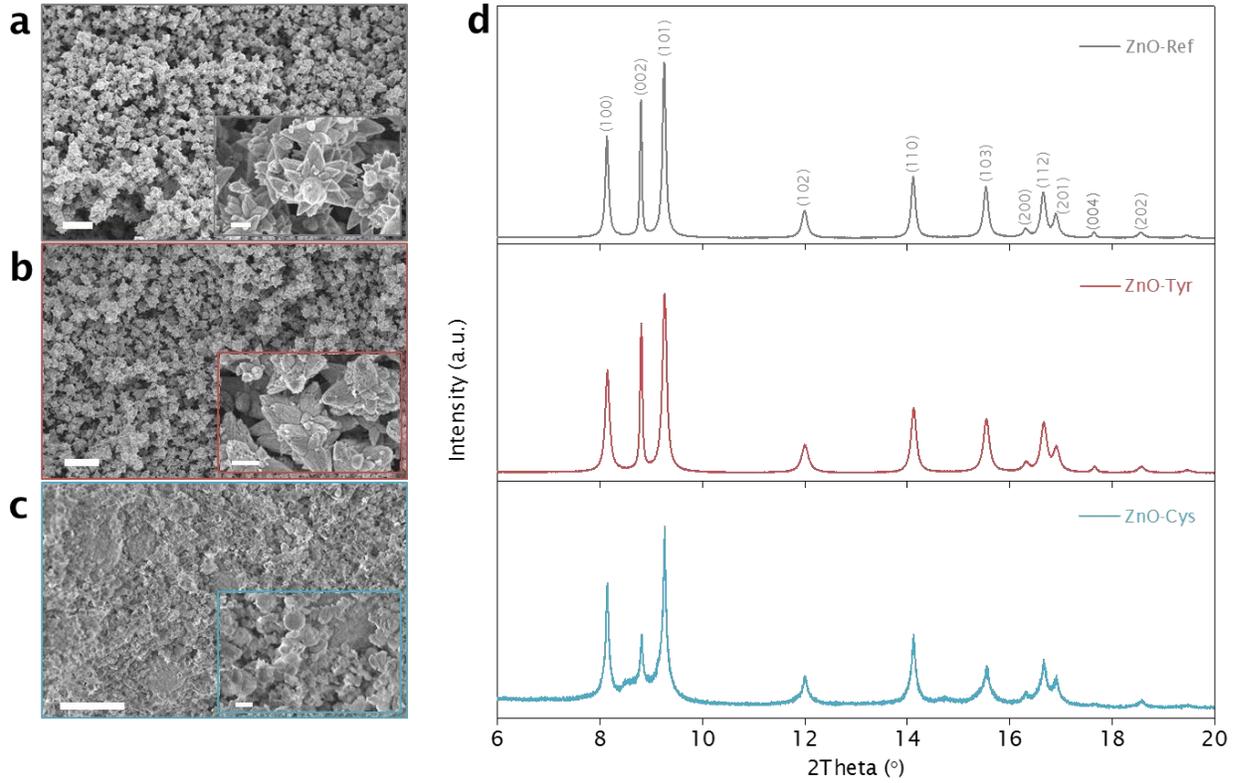

**Figure 1.** SEM micrographs of (a) ZnO-Ref, (b) ZnO-Tyr and (c) ZnO-Cys co-crystals (scale bars, 5 μm). Inset: high magnification micrographs (scale bars, 500 nm). (d) High-resolution XRD patterns of the samples shown in (a-c), respectively.

**Table 1.** Crystal lattice parameters and induced lattice distortions for ZnO-Tyr and ZnO-Cys with respect to ZnO-Ref as determined from Rietveld analysis of the high-resolution XRD patterns.

| Sample | $a, b$ parameters (Å) | relative distortions $a$-axis | $c$ parameter (Å) | relative distortions $c$-axis | $\chi^2$ |
|---|---|---|---|---|---|
| **ZnO-Ref** | 3.250823(2) | - | 5.207636(9) | - | 3.303 |
| **ZnO-Tyr** | 3.254477(9) | 1.124(6)·10⁻³ | 5.214426(1) | 1.304(2)·10⁻³ | 7.626 |
| **ZnO-Cys** | 3.255475(3) | 1.431(6)·10⁻³ | 5.214535(5) | 1.325(2)·10⁻³ | 4.707 |

We first focused on the investigation of the light absorption properties of the ZnO-amino acid co-crystals and ZnO-Ref upon interaction with unpolarized white light and, especially, upon interaction with circularly polarized light. The inset in **Figure 2**a displays the characteristic absorption spectra of the three samples. While the absorption onsets for the ZnO-Tyr and ZnO-Ref samples are quite close, the absorption onset for ZnO-Cys is significantly blue-shifted. The absorption spectra of ZnO-Tyr and, most especially, of ZnO-Cys also show a significant tail at λ≥400 nm that may be ascribed to the formation of crystal defects resulting from amino acid co-crystallization. Even though some differences between the three samples are evident from their absorption spectra, their light absorption differences are especially accentuated upon interaction with circularly polarized light. **Figure 2**a shows the circular dichroism (CD) spectra of ZnO-Ref, ZnO-Tyr and ZnO-Cys, along with the CD spectra of the corresponding amino acids. Both amino acids are characterized by prominent positive CD peaks centered at 208 nm. Cysteine displays a positive CD signal, in agreement with a previous work by Takatori *et al.*,[31] who demonstrated that L-amino acids provide positive CD peaks. In the case of tyrosine, the positive CD signature indicates that the L enantiomer is the dominant one in the DL-tyrosine we have used, *i.e.*, our DL-tyrosine is not a racemic mixture. On the semiconductor side, the CD spectrum of ZnO-Ref is essentially featureless, since there is no chiral molecule to interact with the semiconductor. This is in stark contrast to the CD spectra of ZnO-Tyr and ZnO-Cys. Both samples display strong CD peaks centered at 217 and 215 nm, respectively, ascribed to the intracrystalline amino acids. The polarity of these peaks changes from positive for the pure amino acids to negative in the ZnO-amino acid samples. This polarity inversion can be explained on the basis of the strong sensitivity of CD signals to factors such as the atomic configuration and the surrounding environment.[32] The incorporation and strong interaction of Cys and Tyr with the crystal lattice of the semiconductor is further supported by the chiral ZnO excitons ($\lambda < 400$ nm) and chiral amino acid-induced, defect-related absorption ($\lambda > 400$ nm) that both ZnO-amino acid samples display. The chiral excitons occur as a result of the chirality transfer from the amino acids (tyrosine and cysteine) to the inter-band transitions of the ZnO matrix. Chirality transfer has been reported extensively for II-VI quantum dots stabilized by chiral ligands,[33] including amino acids (for which the intensity of the CD signals is similar to the intensity of the CD signals featured by our ZnO-amino acid co-crystals),[34] but, to the best of our knowledge, this is the first time that this transfer is reported for a semiconductor containing a chiral biomolecule incorporated within its crystal lattice. The strong CD peaks at wavelengths just below the band edge can be attributed to the defects. Defects are most likely located at the interface between the ZnO and the amino acids and, therefore, it is expected that these defect states should be strongly affected by the chirality of the amino acids, thus giving rise to strong CD signals at the wavelengths of defects. On the other hand, the CD peaks at 360-370 nm are likely due to the interband transitions in ZnO. These electronic states are spatially located further from the amino acids and, therefore, it is expected that their CD signals will be weaker. The observed mechanism of chiral transfer from amino acids to the inter-band excitations including the excitons (**Figure 2**b) is a non-trivial process (see additional CD spectra in Figure S5, Supporting Information) and can be due to the motion of electrons and holes within the semiconductor in the presence of the quantum chiral barriers created by the biomolecules, as will be discussed further below. Importantly, despite a variety of physical mechanisms found in bio-conjugated nanocrystal systems,[35-37] this quantum mechanism has not been described so far.

In our previous work[26] a thermal treatment was carried to elucidate the effect of amino acid decomposition on the lattice parameters of the ZnO-amino acid co-crystals. It was demonstrated that the lattice strain of most co-crystals relaxed upon mild thermal annealing in air (300 °C, 90 min). Therefore, in order to complete our chirality study and, therefore, gain further insights into the role of the amino acids on the co-crystals' CD signatures, we tracked their CD signal evolution upon amino acid decomposition at 50°C, 100°C, 150°C, 200°C, 250°C, and 450°C for 90 min. The crystallinity and morphology of all samples remained unchanged in all cases (see Figure S3, Figure S4, Figure S7), while the lattice distortions of the co-crystals are known to relax upon thermal decomposition of the amino acids.[26] Unfortunately, in none of the cases (even for the samples heated at moderate temperatures and still showing the presence of the amino acids, see discussion further below) have we succeeded in getting a CD signal. The most likely reason for this is related to the intrinsically low CD intensity of the original, unheated, samples and the significant increase in noise level for all heated samples. This makes it virtually impossible to distinguish any CD signal stemming from the samples.

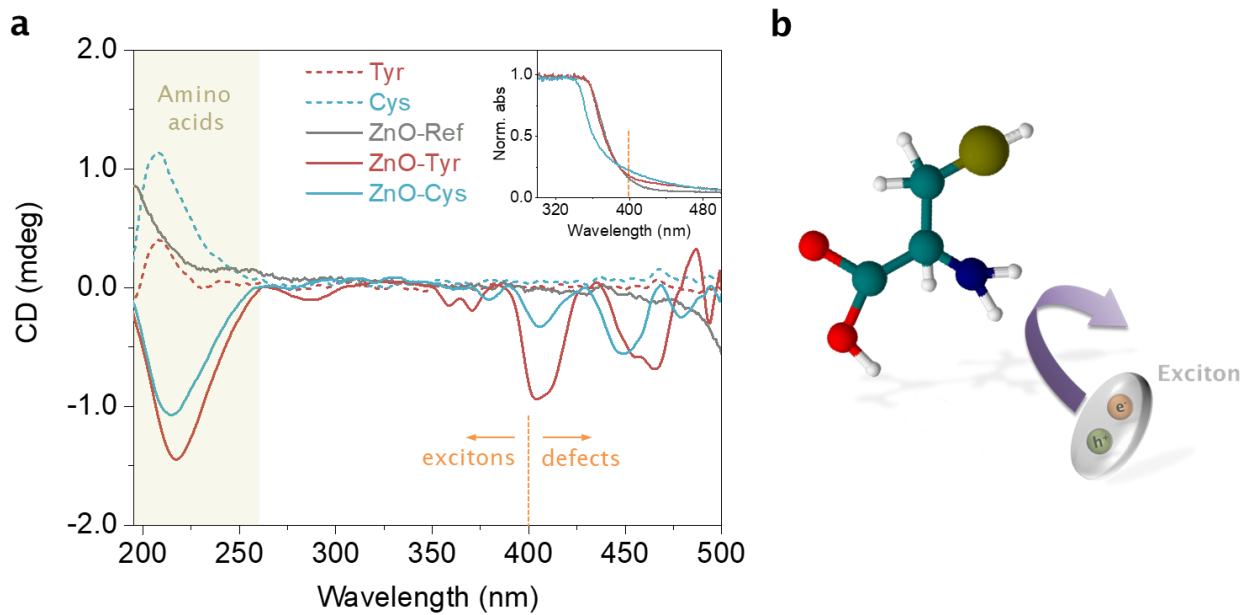

**Figure 2.** (a) Circular dichroism spectra of the pure amino acids Tyr (DL-tyrosine) and Cys (L-cysteine), dashed curves, and of the ZnO-Ref, ZnO-Tyr, and ZnO-Cys samples (solid curves). The CD spectra of the pure amino acids were measured in aqueous solution, while the spectra of the three ZnO samples were recorded in powder form. The yellowish shading highlights the characteristic UV signal of the amino acids in both the pure amino acid samples and co-crystals. Unlike ZnO-Ref, ZnO-Tyr and ZnO-Cys display chiral excitons ($\lambda < 400$ nm) and amino acid-induced, defect-related, chirality signatures ($\lambda > 400$ nm). A delocalized optically-created exciton in ZnO scatters off an amino acid molecule. The motions and the envelope wave function of excitons in ZnO should be strongly affected by the scattering produced by the chiral amino acid biomolecules and, hence, optical CD at the inter-band transition wavelength is expected to become active. Inset: absorption spectra of ZnO-Ref, ZnO-Tyr, and ZnO-Cys as obtained from integrating sphere measurements in powder form. The spectra are normalized at 300 nm (before the absorption onset) for comparison purposes. The 400 nm reference is also shown to illustrate the amino acid-induced, defect-related, absorption at $\lambda>400$ nm. (b) Qualitative picture depicting the interaction

between an exciton in ZnO and a scattering barrier with a chiral shape, an amino acid (here L-cysteine) in our case.

As shown, the intracrystalline amino acids strongly influence the light absorption properties of the ZnO co-crystals. However, their most striking effect is produced within their distinctive light emission. The fluorescence spectra of the three ZnO samples are displayed in **Figure 3**a and show the strong band edge emission blue shift of the ZnO-Ref sample ($E_g$ = 3.26 eV) as a result of amino acid incorporation. The band edge emission energies of ZnO-Tyr and ZnO-Cys peak at 3.28 eV and 3.40 eV, respectively. This means that Tyr induces a 20 meV increase in emission energy, while for the ZnO-Cys co-crystal the increase amounts to 140 meV, a large band emission change, as we will discuss further below. Furthermore, the fluorescent spectra of the three samples in the 1.9 – 3.5 eV range (see Figure S6, Supporting Information) show that not only the band edge emission blue shifts as a result of amino acid co-crystallization, but also the defect emission does blue shift by a similar magnitude.

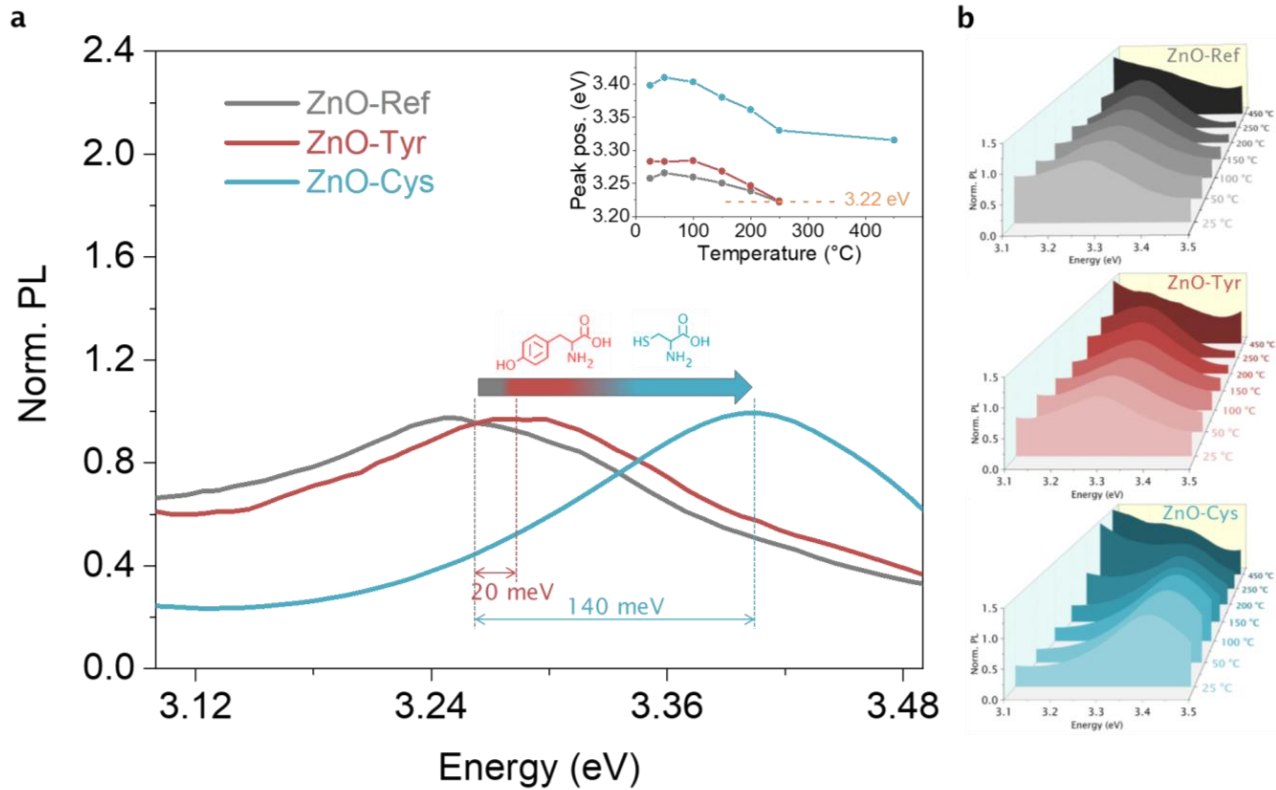

**Figure 3.** (a) Fluorescence spectra of ZnO-Ref, ZnO-Tyr and ZnO-Cys in the band edge emission region ($\lambda_{excitation}$ = 340 nm). The spectra are normalized at the corresponding band edge emission energies to facilitate comparison, namely: 3.26 eV (ZnO-Ref), 3.28 eV (ZnO-Tyr), and 3.40 eV (ZnO-Cys). The color-graded arrow indicates that the band edge emission blue shifts as a consequence of amino acid incorporation within the ZnO crystal lattice. The corresponding shifts are: 20 meV for ZnO-Tyr and 140 meV for ZnO-Cys. Inset: evolution of the room temperature band edge emission energy of ZnO-Ref, ZnO-Tyr, and ZnO-Cys as a function of the temperatures used for the 90 min thermal treatment. The peak positions were determined from the spectra shown in (**b**) using peak fitting. The peak position of the room temperature band edge emission of ZnO-Ref treated at 250°C ($E_g$ = 3.22 eV) is plotted as reference. (b) Spectral evolution of the room temperature band edge emission of ZnO-Ref (top), ZnO-Tyr (middle), and ZnO-Cys (bottom) after thermal treatment of 6 independent powder aliquots from each sample at 50 °C, 100 °C, 150 °C,

200 °C, 250 °C and 450 °C for 90 min. Analogous plots in 2D are provided in the Supporting Information, Figure S8). The samples referred to as '25°C' are the as-synthesized (unheated) reference samples. For comparison purposes all spectra have been normalized at the band edge emission value of the corresponding 25°C sample. Note that the ZnO-Ref and ZnO-Tyr samples heated at 450°C do not show a distinguishable band edge emission peak. Also note that all fluorescence spectra shown in Figure 3 do not correspond to circularly polarized light emission.

With the aim of gaining insights into the actual role of the amino acids on the blue-shifted band edge emission of the co-crystals with respect to the ZnO-ref, we also investigated the effect of a 90 min thermal treatment at 50°C, 100°C, 150°C, 200°C, 250°C, and 450°C on the room temperature band edge emission of ZnO-Ref, ZnO-Tyr, and ZnO-Cys (**Figure 3**b). We restrict ourselves here to the discussion of the main outcomes of this investigation on the basis of the band edge emission shifts that are summarized in **Figure 3**a, inset (the evolution of the defect emission upon annealing is presented and discussed in Figure S9). Overall, the room temperature band edge emission of ZnO-Ref gradually red-shifts with increasing temperature, reaching an overall red-shift of ~ 36 meV at 250°C ($E_{g\,(ZnO\text{-}Ref,\,250°C)} = 3.22$ eV) with respect to the non-heated sample. This shift is mainly ascribed to the gradual removal of water molecules trapped within the ZnO lattice during the synthesis process (see mass spectra in Figure S10). The room temperature band edge emission of ZnO-Tyr also red-shifts with increasing annealing temperature. However, after 90 min annealing at 250°C, its band edge emission red-shifts up to 3.22 eV, which is the band edge emission energy reached by the ZnO-Ref after thermal treatment at the very same temperature. We attribute this overall ~ 60 meV band edge emission red shift for ZnO-Tyr with respect to the unheated sample to the thermal decomposition of Tyr under those mild thermal annealing conditions.[38] Unlike for ZnO-Tyr, the band edge emission peak of the ZnO-Cys sample heated at 250°C (3.33 eV) does not fully shift back to the emission energy of the ZnO-Ref at the same temperature (3.22 eV). Actually, further heating at 450°C leads to a band edge emission peak centered at 3.32 eV (*i.e.*, to a ~ 83 meV red-shift with respect to the unheated sample), which is still far away from the 3.22 eV of the ZnO-Ref microcrystals. The fact that the band edge emission of the ZnO-Cys sample cannot shift completely back to the value of the ZnO-Ref is probably due to the higher thermal stability of cysteine in the ZnO lattice, which is likely enhanced by the strong interaction between the S and Zn atoms.[39] The mass spectra of ZnO-Cys (Figure S10) indicate a significant release of OH, $H_2O$, $CO_2$, and $NH_3$ species between 250°C and 400°C upon cysteine decomposition and an onset for further $CO_2$ release above ~450°C. This latter result further confirms that, differently from the ZnO-Tyr sample, intracrystalline cysteine may still be present in the ZnO-Cys crystal lattice, even after annealing at 450°C and, therefore, the band edge emission of the co-crystal remains red-shifted with respect to the ZnO-Ref. Overall, these thermal annealing experiments prove that there is a clear correlation between the extent of amino acid decomposition and the extent of band edge emission blue shift in the co-crystals: when all amino acids decompose fully (the case of ZnO-Tyr after heating at 250 °C for 90 min), the band edge emission energy of the co-crystals is analogous to that of the ZnO-Ref. However, when the amino acids do not decompose fully (the case of ZnO-Cys), the band edge emission energy of the co-crystals remains blue-shifted with respect to the ZnO-Ref.

On the basis of the results presented so far, it may appear at first sight reasonable to ascribe the larger band edge emission energy of the co-crystals to the lattice distortions resulting from the amino acid-induced tensile strain.[26] Also the fact that the room temperature band edge emission of ZnO-Tyr red-shifts back to that of ZnO-Ref upon heating at 250°C, but that the band edge emission of ZnO-Cys does not (**Figure 3**a, inset), appears to correlate quite well with the almost full lattice strain relaxation reported for ZnO-Tyr, but not for ZnO-Cys, under similar thermal treatment conditions.[26] However, a careful examination of our data reveals that the assignment of those band edge emission changes to the amino acid-induced lattice distortions is not that straightforward. Previous work by Wei *et al.* demonstrated that ZnO nanowires with an average diameter of 760 nm (commensurate with the dimensions of our microcrystals) undergo a 59 meV near-band-edge emission red-shift under a tensile strain of 1.7%.[4] They showed a similar behavior for thinner nanowires, reporting up to a 110 meV red-shift for 100 nm thick nanowires under a 7.3% tensile strength. In our case, and regardless of the amino acid incorporated, the band edge emission systematically blue-shifts; and this occurs in spite of the tensile lattice strain induced by the intracrystalline amino acids. This is in stark contrast to the red-shifts reported by Wei *et al.*[4] under tensile strain. Furthermore, if we set the focus on our most extreme sample, ZnO-Cys, its band edge emission is 140 meV blue-shifted with respect to the ZnO-Ref and its average lattice strain (*a* and *c* axis-averaged) is 0.1378%. Even though this tensile strength is *ca.* 10-fold lower than the 1.7% needed to induce a 59 meV red-shift in the 760 nm nanowires, the band edge emission blue shift for the ZnO-Cys is *ca.* 2.4-fold higher. Therefore, the tensile strain alone can explain neither our observed emission energy blue-shifts, nor the high values of those shifts for the amino acid-containing ZnO samples. Also, since the crystallite size of ZnO is significantly larger than 2.34 nm (exciton Bohr radius of ZnO) in all samples (ZnO-Ref, ZnO-Tyr, and ZnO-Cys – see Table S1, Supporting Information)[30] conventional quantum size effects cannot be the reason for the higher band edge emission energy of the ZnO-amino acid co-crystals.

Our high-resolution XRD data indicate that the ZnO-amino acid co-crystals can be envisioned as a conventional three-dimensional (3D) ZnO crystal lattice subjected to an anisotropic tensile lattice strain along the *a* and *c* axes. Furthermore, from XRD and WDS analysis we know that the amino acids are indeed incorporated into the ZnO crystal lattice and, in turn, they are responsible for the induced strains. Hence, in this scenario the ZnO-amino acid co-crystals could be visualized as a ZnO-amino acid superlattice whose strain and periodicity ($a_{AA}$) are determined by the entrapped amino acids (see **Figure 4**a and **Figure 4**b). In this organic-inorganic 3D superlattice the amino acids may serve as 3D potential barriers (obstacles) that can indeed change the energy of electrons and holes, as schematically depicted for the ZnO-Cys sample in **Figure 4**c. In order to support our hypothesis for the origin of the band edge emission blue shifts in our ZnO-amino acid co-crystals, we propose here a simple quantum 3D model, which nevertheless should give reasonable estimates for the blue shifts of the excitons. For simplicity we assumed that the amino acids are spheres providing increased potentials (1.5 eV) for electrons and holes (see **Figure 4**c, for the ZnO-Cys sample) and we arranged such barriers in a periodic 3D array, as sketched in **Figure 4**d. We are aware that this latter assumption is a simplification for modeling purposes. In reality, the amino acids may not be perfectly periodically distributed within the ZnO matrix. Instead, they may likely feature a statistical, quasi-periodic, distribution. Even though a technique such as SAXS could provide information about the actual amino acid distribution, this type of characterization is not feasible with our samples due to the very low content of amino acids within the co-crystals (~1%) and their also very low density contrast.[ii] Alternatively, we have performed ellipsometry

---

[ii] Additional note: if we had areas of agglomerated amino acids rather than a distribution of those amino acid crystals within the ZnO lattice, we would expect a splitting of the diffraction peaks in the co-crystals, that is, we would have diffraction peaks from areas where the amino acids are distributed in the lattice (strained) and from areas lacking the amino acids. However, our XRD data show that the entire lattice is strained. Also, in an earlier work Kim *et al*. (Kim, Y.-Y.;

measurements on our samples in order to determine the porosity level that the co-crystalized amino acids may induce and, hence, to be able to derive a realistic model accounting for the actual amino acid distribution within the ZnO. However, no reliable ellipsometry data were obtained neither from our powder ZnO-amino acid samples nor from their corresponding thin films.[iii] Therefore, for modeling simplicity we have assumed a periodic 3D distribution of amino acids within the ZnO lattice. In our model the heights of the amino acid barriers reflect the band-gap energy of ZnO and the HOMO and LUMO energies of the amino acids. We estimated the period of the amino acid barriers, *i.e.*, the superlattice period $a_{AA}$ (namely, 1.941 nm for ZnO-Tyr and 0.934 nm for ZnO-Cys) from the estimated amino acid diameters ($d_{AA}$) and the particular amino acid concentration in each of the synthesized ZnO-Tyr and ZnO-Cys samples (obtained from WDS analysis). Details of the model and particular parameters can be found in the Experimental Section

---

Ganesan, K.; Yang, P.; Kulak, A. N.; Borukhin, S.; Pechook, S.; Ribeiro, L.; Kröger, R.; Eichhorn, S. J.; Armes, S. P.; Pokroy, B.; Meldrum, F. C., An artificial biomineral formed by incorporation of copolymer micelles in calcite crystals. *Nat Mater* **2011,** *10* (11), 890-896) showed that calcite can be co-crystallized with diblock co-polymer micelles. Those co-crystals feature a much higher incorporation of the organic component than our ZnO-amino acid co-crystals, but analogous XRD peak shifts. Nevertheless, in that case it was proven by NMR that there is no agglomeration of the biomacromolecules, which further supports that our amino acids are statistically distributed within the ZnO lattice.

[iii] Thin film formation with the ZnO-amino acid co-crystal samples is challenging. Upon redispersion in several solvents (water, ethanol, isopropanol, etc.) we observed that the co-crystallized amino acids systematically solubilized and diffused from the ZnO crystal lattice into solution. This was confirmed by the fact that the band edge emission of the ZnO-amino acid co-crystals blue shifted back to the band edge emission of the ZnO-ref after solvent redispersion, thus confirming the solubilization and out-diffusion of the originally co-crystallized amino acids. For this reason, so far it has not been possible to find a suitable thin film preparation method that enables a reliable ellipsometry analysis on our samples. Alternative thin film preparation routes will be pursued in the future.

and in the quantum modeling section in the Supporting Information. The most important results from our quantum model are that the spectra of carriers (electrons and holes) in the ZnO-amino acid co-crystals form new sub-bands in the conduction and valence bands (see Figure S11) and that, as a consequence of the energy shifts of those sub-bands, the inter-band gap in the ZnO-amino acid co-crystals exhibits a blue shift with respect to the ZnO-Ref. This is illustrated in **Figure 4c** for the ZnO-Cys sample. The lowest energy of the new sub-band for the electrons in ZnO-Cys is shifted 77 meV upwards (higher energy) with respect to the reference ZnO; while for the holes the shift amounts to 70 meV downwards (lower energy). These shifts result in an overall increase in emission energy of $\Delta E_g$=147.0±1.7 meV for ZnO-Cys, while for the ZnO-Tyr sample our model predicts an energy increase of $\Delta E_g$=21.0±0.6 meV. In both cases, and considering the simplicity of the model, which, for instance, does not take into account neither the possibility of amino acid clustering within the ZnO lattice nor the presence of the $Zn_5(OH)_8(NO_3)_2$ impurity (< 5% wt.) in the ZnO-Cys sample,[iv] the calculated inter-band energies are in good agreement with the experimentally-observed blue shifts, which are 140 meV and 20 meV for ZnO-Cys and ZnO-Tyr, respectively.

---

[iv] If the possibility of amino acid clustering in both ZnO-Tyr and ZnO-Cys were taken into account in our model, as well as the amount of $Zn_5(OH)_8(NO_3)_2$ impurity present in the ZnO-Cys sample, the periodic potential of both co-crystals would feature superlattice periods ($a_{AA}$) larger than the ones estimated herein (namely, 1.941 nm for ZnO-Tyr and 0.934 nm for ZnO-Cys). Therefore, the band edge emission energies predicted would be smaller than the ones estimated with the current model (21 meV for ZnO-Tyr and 147 meV for ZnO-Cys).

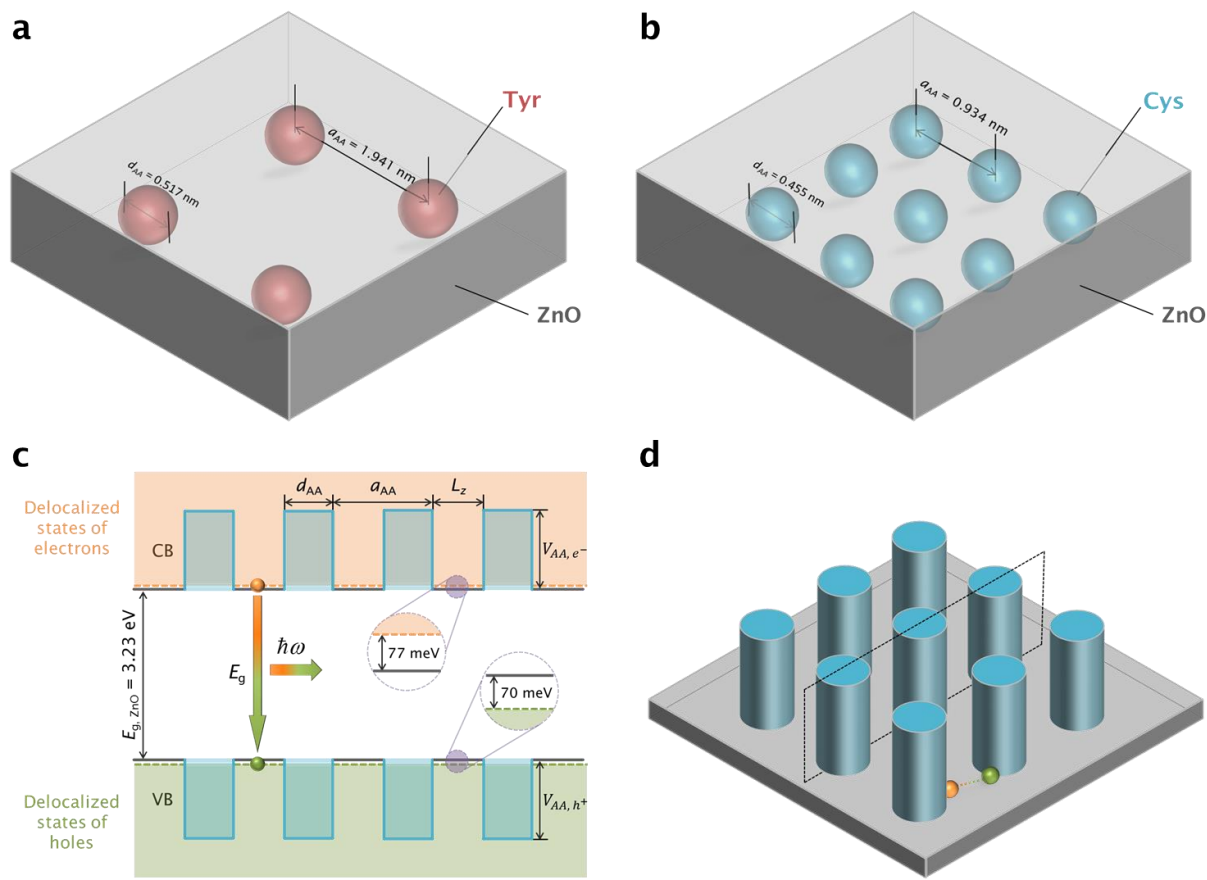

**Figure 4.** Schematic 3D representation of an idealized ZnO crystal containing co-crystalized (a) tyrosine or (b) cysteine. Both drawings are scale representations and include the specific superlattice parameters ($a_{AA}$: superlattice period, $L_z$: amino acid-to-amino acid surface-to-surface gap) as determined from the amino acid loadings of each sample. The estimated diameter of each amino acid ($d_{AA}$) is also indicated. (c) Band structure of the super-periodic ZnO-amino acid co-crystals. Due to the interaction with the periodic array of amino acid barriers the wave functions of the carriers in ZnO form new sub-bands (see Figure S11, Supporting Information) and the lower energy states in the conduction and valence bands exhibit the energy shifts. The resulting inter-band transition for the PL process receives a quantum blue shift that is observed experimentally. The energy shifts depicted in panel (c) (77 meV for electrons, 70 meV for holes) correspond to the ZnO-Cys case, where the calculated overall energy change with respect to ZnO is 147 meV. (d)

Illustrative 3D depiction of the periodic potential landscape provided by cysteine in the ZnO-Cys co-crystals. The vertical plane (dashed line) represents the 2D cut that is shown in panel (c). The figure illustrates that delocalized electrons and holes in the super-periodic ZnO-amino acid co-crystals move in the presence of the amino acid potential barriers.

Additionally, we also examined the effect of the superlattice period on the calculated energy shifts of the ZnO-amino acid co-crystals from a physics perspective. The obtained results are summarized in **Figure 5** for ZnO-Cys and in Figure S12, for ZnO-Tyr. All calculated energy values at each superlattice period were obtained at a constant ZnO-to-amino acid volume ratio, namely $V_{AA}$ / $V_{ZnO}$ = 0.064 for ZnO-Cys and $V_{AA}$ / $V_{ZnO}$ = 0.010 for ZnO-Tyr, since both ratios reflect the actual amino acid loading in the corresponding experimental ZnO-amino acid samples. Therefore, with decreasing the superlattice period, the number of amino acid molecules increases so that the $V_{AA}$ / $V_{ZnO}$ volume ratio is kept constant. The results obtained confirm the gradual energy increase for electrons and holes and, in turn, for the inter-band energy with decreasing superlattice period (see tabulated values for selected $a_{AA}$ values).

Our modeling results support the experimental thermal annealing results discussed above, namely, that there is a clear correlation between the extent of amino acid decomposition and the extent of band edge emission blue shift in the co-crystals. According to our quantum model, the gradual decomposition of the amino acids leads to a gradually increasing amino acid-to-amino acid spacing in the co-crystals and, thereby, to an increased superlattice period ($a_{AA}$) that, ultimately, leads to a decreased band edge emission energy (see curves in Figure 5 and Figure S12), in agreement with our experimental results (Figure 3). Furthermore, our modeling results also explain why the band edge emission energy of the ZnO-Cys sample (3.40 eV) is larger than the band edge emission energy of the ZnO-Tyr one (3.28 eV). At first sight one could ascribe this to the minor impurity phase ($Zn_5(OH)_8(NO_3)_2$, < 5% wt.) present in the ZnO-Cys sample, but not in the ZnO-Tyr one. The amount of this impurity phase does not change even after air annealing at 300°C for 2 hours (Figure S4). However, we have shown that the band edge emission energy of ZnO-Cys does decrease under those annealing conditions. Therefore, our results rule out any influence of this impurity phase on the higher band edge emission of ZnO-Cys with respect to ZnO-Tyr. On the other hand, one could also hypothesize that morphological differences between the ZnO-Cys and the ZnO-Tyr co-crystals could play a role in the higher band edge emission of the former. Here, the results depicted in Figure S7 (crystallinity and morphology of ZnO-Ref, ZnO-Tyr, and ZnO-Cys after thermal treatment) illustrate that the morphology of both ZnO-Cys and ZnO-Tyr remains unchanged upon heating up to 450°C. However, we know that such an annealing leads to a gradual decrease in band edge emission energy for both samples (see results in Figure 3 and associated discussion). Therefore, morphological differences cannot be the reason for the higher band edge emission energy of ZnO-Cys. Indeed, the reason for the higher band edge emission energy of the ZnO-Cys sample *vs*. the ZnO-Tyr sample is its higher amino acid incorporation (% atomic N: 1.70 ± 0.03 for ZnO-Cys *vs*. 0.19 ± 0.03 for ZnO-Tyr). According to our proposed quantum model, a higher amount of intracrystalline amino acids leads to a shorter amino acid-to-amino acid spacing and, thereby, to a smaller superlattice period ($a_{AA}$(ZnO-Cys) = 0.934 nm *vs*. $a_{AA}$(ZnO-Tyr) = 1.941 nm – see Figure 4a and 4b). Therefore, it is the larger superlattice period of ZnO-Cys what explains the higher band edge emission energy of this sample *vs.* the ZnO-Tyr one.

Overall, the good qualitative agreement of the experimental inter-band energy shifts of ZnO-Cys and ZnO-Tyr with our simple model highlights the importance of quantum confinement effects in the ZnO-amino acid co-crystals and, hence, provides an understanding on the origin of the band gap tuning observed in the ZnO-amino acid samples.

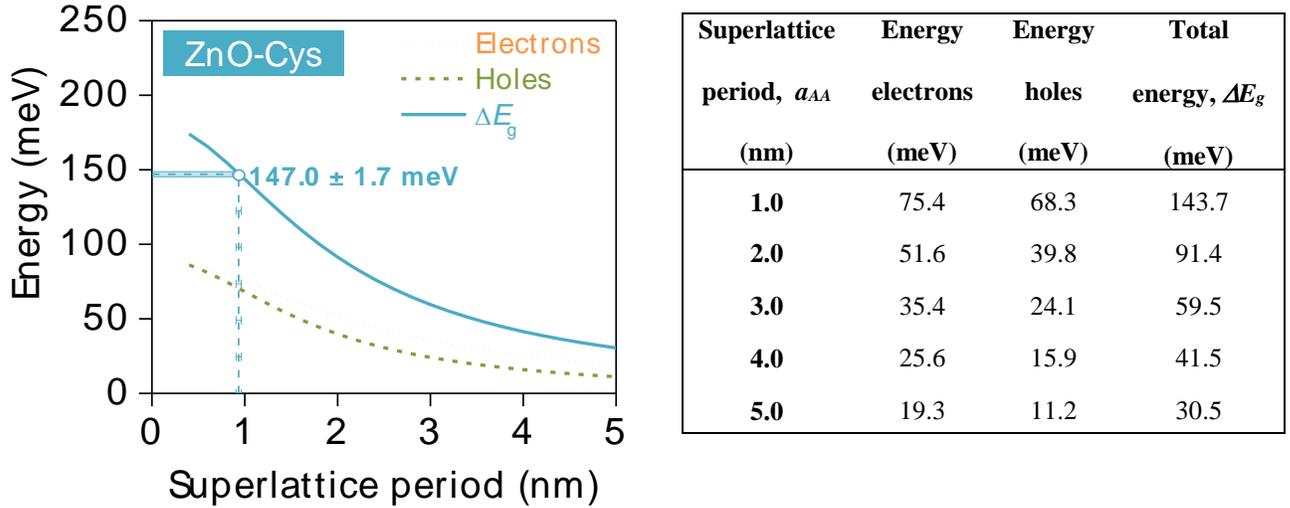

| Superlattice period, $a_{AA}$ (nm) | Energy electrons (meV) | Energy holes (meV) | Total energy, $\Delta E_g$ (meV) |
|---|---|---|---|
| 1.0 | 75.4 | 68.3 | 143.7 |
| 2.0 | 51.6 | 39.8 | 91.4 |
| 3.0 | 35.4 | 24.1 | 59.5 |
| 4.0 | 25.6 | 15.9 | 41.5 |
| 5.0 | 19.3 | 11.2 | 30.5 |

**Figure 5** Calculated shifts for the lowest-energy states with zero momenta for the conduction and valence bands (electrons and holes, respectively) of ZnO-Cys co-crystals as a function of the superlattice period. The calculated shifts were obtained at a constant ZnO-to-amino acid volume ratio, namely $V_{AA} / V_{ZnO} = 0.064$, *i.e.*, the ratio corresponding to the actual Cys loading of our experimental ZnO-Cys sample. This volume ratio conservation implies that the number of Cys molecules (potential barriers) increases as the superlattice period decreases. The calculations show that with decreasing the supperlattice period the total inter-band energy ($\Delta E_g$) increases. The $\Delta E_g$ = 147.0 ± 1.7 meV marked in the graph corresponds to our experimental ZnO-Cys sample, for which the superlattice period is $a_{AA} = 0.934 \pm 0.028$ nm. The uncertainty of the calculated inter-band shifts derives from the uncertainty in the superlattice period, which arises from the uncertainty of the WDS measurements used to determine the Cys loading in the ZnO-Cys sample. Given for reference, the table contains the calculated values (energy of electrons and holes and total inter-band energy) of ZnO-Cys for specific superlattice periods.

In this work we have demonstrated that the light-matter interactions in ZnO-amino acid co-crystals are dramatically affected by the presence of the co-crystallized biomolecules. We have disentangled the role played by the amino acids on the band gap modulation of the semiconductor. On the basis of the good qualitative agreement between the experimental and calculated increases in band edge emission energy we propose that the amino acids may serve as quasi-super-periodic 3D potential barriers for electrons and holes in ZnO. In this scenario, the carrier wave functions in ZnO form new sub-bands that result in an energy increase between the lower energy states of the conduction and valence bands and, hence, in a quantum PL blue shift. This result, together with the proven chirality transfer from the intracrystalline amino acids to the inter-band excitations of ZnO, reveal that biomolecule co-crystallization can be harvested as a novel and truly biomimetic approach to prompt chiral quantum confinement effects in semiconductors. Hence, bio-inspired biomolecule-semiconductor co-crystals could serve as photonic materials for chiral quantum optics.[40]

ASSOCIATED CONTENT

The Supporting Information is available free of charge on the ACS Publications website. Experimental section, circular dichroism and fluorescence spectra of all samples, XRD patterns and SEM micrographs of all samples at different annealing temperatures, mass spectra, quantum modeling details, calculated spectra for electrons and holes in ZnO-Cys, and calculated effect of the superlattice period on the emission energy of ZnO-Tyr. (PDF)

AUTHOR INFORMATION

**Corresponding Author**

*e-mail: jessica.rodriguez@lmu.de, govorov@ohio.edu, bpokroy@tx.technion.ac.il

**Author Contributions**

All authors contributed to the design of the experiments and the interpretation of the results. M.A.H.B., M.L., V.B., P.D., A.J.B., I.P., and A.N.F carried out the experiments. X.T.K. and A.O.G. implemented the quantum modeling. J.R.F. wrote the manuscript with input and comments from all authors.

**Notes**

The authors declare no competing financial interest.

ACKNOWLEDGMENTS

We thank support by the Bavarian State Ministry of Science, Research, and Arts through the grant "Solar Technologies go Hybrid" (SolTech). A.O.G was supported by the Volkswagen Foundation. X.-T. K. was supported by the oversea postdoc program of Institute of Fundamental and Frontier Sciences, University of Electronic Science and Technology of China. B.P. acknowledges the funding received from the European Research Council under the European Union's Seventh Framework Program (FP/2007-2013)/ERC Grant Agreement (no. 336077). We also acknowledge ID22 of the ESRF and Dr. Andy Fitch for support with these experiments. We also thank Prof.

Tim Liedl and Dr. Eva Maria Roller for granting access to the CD instrumentation and technical support.